\documentclass[acmlarge,10pt]{acmart}

\AtBeginDocument{%
	\providecommand\BibTeX{{%
			\normalfont B\kern-0.5em{\scshape i\kern-0.25em b}\kern-0.8em\TeX}}}

\renewcommand\footnotetextcopyrightpermission[1]{} 
\usepackage[utf8]{inputenc}
\usepackage{booktabs} 
\usepackage{tabularx}
\usepackage[ruled]{algorithm2e} 

\usepackage{xcolor}
\usepackage{verbatim}

\usepackage{colortbl}

\usepackage{units}
\usepackage{makecell}

\usepackage{graphicx}

\definecolor{grey1}{rgb}{0.5,0.5,0.5}
\definecolor{grey2}{rgb}{0.9,0.9,0.9}

\usepackage{pifont}
\newcommand{\xmark}{\ding{55}}%

\SetAlFnt{\small}
\SetAlCapFnt{\small}
\SetAlCapNameFnt{\small}
\SetAlCapHSkip{0pt}
\IncMargin{-\parindent}

\newcommand{\gru}{\textsc{gru4rec}\xspace}
\newcommand{\grutwo}{\textsc{gru4rec2}\xspace}
\newcommand{\igru}{\textsc{gru4rec+}\xspace}
\newcommand{\knn}{\textsc{s-knn}\xspace}

\newcommand{\sr}{\textsc{sr}\xspace}
\newcommand{\ar}{\textsc{ar}\xspace}
\newcommand{\mc}{\textsc{mc}\xspace}
\newcommand{\iknn}{\textsc{iknn}\xspace}
\newcommand{\sknn}{\textsc{sknn}\xspace}
\newcommand{\vsknn}{\textsc{v-sknn}\xspace}

\newcommand{\smf}{\textsc{smf}\xspace}
\newcommand{\fpmc}{\textsc{fpmc}\xspace}
\newcommand{\bpr}{\textsc{bpr-mf}\xspace}
\newcommand{\fism}{\textsc{fism}\xspace}
\newcommand{\fossil}{\textsc{fossil}\xspace}

\newcommand{\narm}{\textsc{narm}\xspace}
\newcommand{\stamp}{\textsc{stamp}\xspace}
\newcommand{\nnet}{\textsc{nextitnet}\xspace}
\newcommand{\csrm}{\textsc{csrm}\xspace}
\newcommand{\sgnn}{\textsc{sr}\mbox{-}\textsc{gnn}\xspace}

\newcommand{\ct}{\textsc{ct}\xspace}
\newcommand{\stan}{\textsc{stan}\xspace}
\newcommand{\vstan}{\textsc{vstan}\xspace}

\newcommand{\rsc}{RSC15\xspace}
\newcommand{\retailrocket}{RETAIL\xspace}
\newcommand{\zalando}{ZALANDO\xspace}

\newcommand{\aotm}{AOTM\xspace}
\newcommand{\tmusic}{30MU\xspace}
\newcommand{\nowplaying}{NOWP\xspace}
\newcommand{\diginetica}{DIGI\xspace}
\newcommand{\etracks}{8TRACKS\xspace}

\newcommand{\best}[1]{\textbf{#1}}
\newcommand{\sign}[1]{$*$\textbf{#1}}

\newcommand{\scnd}[1]{\underline{#1}}

\newcommand{\spotify}{\textsc{spotify}\xspace}
\newcommand{\cagh}{\textsc{cagh}\xspace}

\definecolor{color1}{RGB}{43,131,186}
\definecolor{color2}{RGB}{171,221,164}
\definecolor{color3}{RGB}{215,25,28}
\definecolor{color4}{RGB}{253,174,97}

\newcommand{\td}[1]{%
	\IfSubStr{#1}{S}{%
		*\StrBehind{#1}{S}[\tmp]\StrLeft{\tmp}{6}[\tmpp]\num[math-rm=\mathbf]{\tmpp}%
	}{%
		\IfSubStr{#1}{F}{%
			\StrBehind{#1}{F}[\tmp]\StrLeft{\tmp}{6}[\tmpp]\num[math-rm=\mathbf]{\tmpp}%
		}{%
			\StrLeft{#1}{6}[\tmp]\num{\tmp}%
		}%
	}%
}%

\newcommand{\tdz}[1]{%
	\sisetup{%
		round-mode          	= places, 
		round-precision     	= 0, 
		detect-weight			= true,
		detect-inline-weight	= math
	}%
	\IfSubStr{#1}{S}{%
		*\StrBehind{#1}{S}[\tmp]\StrLeft{\tmp}{6}[\tmpp]\num[math-rm=\mathbf]{\tmpp}%
	}{%
		\IfSubStr{#1}{F}{%
			\StrBehind{#1}{F}[\tmp]\StrLeft{\tmp}{6}[\tmpp]\num[math-rm=\mathbf]{\tmpp}%
		}{%
			\StrLeft{#1}{6}[\tmp]\num{\tmp}%
		}%
	}%
}%
\newcommand{\tdn}[1]{%
	\sisetup{%
		round-mode          	= places, 
		round-precision     	= 2, 
		detect-weight			= true,
		detect-inline-weight	= math
	}%
	\IfSubStr{#1}{S}{%
		*\StrBehind{#1}{S}[\tmp]\StrLeft{\tmp}{6}[\tmpp]\num[math-rm=\mathbf]{\tmpp}%
	}{%
		\IfSubStr{#1}{F}{%
			\StrBehind{#1}{F}[\tmp]\StrLeft{\tmp}{6}[\tmpp]\num[math-rm=\mathbf]{\tmpp}%
		}{%
			\StrLeft{#1}{6}[\tmp]\num{\tmp}%
		}%
	}%
}%

\begin{document}

\title[Empirical Analysis of Session-Based Recommendation Algorithms]{Empirical Analysis of Session-Based Recommendation Algorithms}
\subtitle{A Comparison of Neural and Non-Neural Approaches}


\author{Malte Ludewig}
\affiliation{%
	\institution{TU Dortmund}
	\country{Germany}}
\email{malte.ludewig@tu-dortmund.de}

\author{Noemi Mauro }
\affiliation{%
	\institution{University of Torino}
	\country{Italy}}
\email{noemi.mauro@unito.it}

\author{Sara Latifi}
\affiliation{%
	\institution{University of Klagenfurt}
\country{Austria}}
\email{sara.latifi@aau.at}

\author{Dietmar Jannach}
\affiliation{%
	\institution{University of Klagenfurt}
	\country{Austria}}
\email{dietmar.jannach@aau.at}

\date{}
\fancyfoot{}
\renewcommand{\shortauthors}{Ludewig et al.}

\begin{abstract}
	Recommender systems are tools that support online users by pointing them to potential items of interest in situations of information overload. In recent years, the class of session-based recommendation algorithms received more attention in the research literature. These algorithms base their recommendations solely on the observed interactions with the user in an ongoing session and do not require the existence of long-term preference profiles. Most recently, a number of deep learning based (``neural'') approaches to session-based recommendations were proposed. However, previous research indicates that today's complex neural recommendation methods are not always better than comparably simple algorithms in terms of prediction accuracy.
	
	With this work, our goal is to shed light on the state-of-the-art in the area of session-based recommendation and on the progress that is made with neural approaches. For this purpose, we compare twelve algorithmic approaches, among them six recent neural methods, under identical conditions on various datasets. We find that the progress in terms of prediction accuracy that is achieved with neural methods is still limited. In most cases, our experiments show that simple heuristic methods based on nearest-neighbors schemes are preferable over conceptually and computationally more complex methods. Observations from a user study furthermore indicate that recommendations based on heuristic methods were also well accepted by the study participants. To support future progress and reproducibility in this area, we publicly share the \textsc{\small{session-rec}} evaluation framework that was used in our research.\footnote{This work combines and significantly extends our own previous work published in \citep{ludewigjannach2019radio} and \citep{LudewigMauro2019}. This paper is void of plagiarism or self-plagiarism as defined by the Committee on Publication Ethics and Springer Guidelines.}
	
\end{abstract}


\maketitle
\thispagestyle{empty}


\section{Introduction}
\label{sec:introduction}
Recommender systems (RS) are software applications that help users in situations of information overload and they have become a common feature on many modern online services. Collaborative filtering (CF) techniques, which are based on behavioral data collected from larger user communities, are among the most successful technical approaches in practice. Historically, these approaches mostly rely on the assumption that information about longer-term preferences of the individual users are available, e.g., in the form of a user-item rating matrix \citep{Resnick:1994:GOA:192844.192905}. In many real-world applications, however, such longer-term information is often not available, because users are not logged in or because they are first-time users. In such cases, techniques that leverage behavioral patterns in a community can still be applied \citep{JannachZankerCF2018}. The difference is that instead of the long-term preference profiles only the observed interactions with the user in the ongoing session can be used to adapt the recommendations to the assumed needs, preferences, or intents of the user. Such a setting is usually termed a \emph{session-based recommendation} problem \cite{QuadranaetalCSUR2018}.

Interestingly, research on session-based recommendation was very scarce for many years despite the high practical relevance of the problem setting. Only in recent years, we can observe an increased interest in the topic in academia \cite{DBLP:journals/corr/abs-1902-04864}, which is at least partially caused by the recent availability of public datasets in particular from the e-commerce domain. This increased interest in session-based recommendations coincides with the recent boom of deep learning (neural) methods in various application areas. Accordingly, it is not surprising that several neural session-based recommendation approaches were proposed in recent years, with \gru being one of the pioneering and most cited works in this context \cite{Hidasi2016GRU}.

From the perspective of the evaluation of session-based algorithms, the research community---at the time when the first neural techniques were proposed---had not yet established a level of maturity as is the case for problem setups that are based on the traditional user-item rating matrix. This led to challenges that concerned both the question what represents the state-of-the-art in terms of algorithms and the question of the evaluation protocol when time-ordered user interaction logs are the input instead of a rating matrix. Partly due to this unclear situation, it soon turned out that in some cases comparably simple non-neural techniques, in particular ones based on nearest-neighbors approaches, can lead to very competitive or even better results than neural techniques \cite{JannachLudewig2017RecSys,Ludewig2018}.
Besides being competitive in terms of accuracy, such more simple approaches often have the advantage that their recommendations are more transparent and can more easily be explained to the users. Furthermore, these simpler methods can often be updated online when new data becomes available, without requiring expensive model retraining.

However, during the last few years after the publication of \gru, we have mostly observed new proposals in the area of complex models. With this work, our aim is to assess the progress that was made in the last few years in a reproducible way. To that purpose, we have conducted an extensive set of experiments in which we compared twelve session-based recommendation techniques under identical conditions on a number of datasets. Among the examined techniques, there are six recent neural approaches, which were published at highly-ranked publication outlets such as KDD, AAAI, or SIGIR after the publication of the first version of \gru in 2015.\footnote{Compared to our previous work presented in \cite{Ludewig2018} and \cite{LudewigMauro2019}, our present analysis includes considerably more recent deep learning techniques and baseline approaches. We also provide the outcomes of additional measurements regarding the scalability and stability of different algorithms. Finally, we also contrast the outcomes of the offline experiments with the findings obtained in a user study \cite{ludewigjannach2019radio}.}

The main outcome of our offline experiments is that the progress that is achieved with neural approaches to session-based recommendation is still limited. In most experiment configurations, one of the simple techniques outperforms all the neural approaches. In some cases, we could also not confirm that a more recently proposed neural method consistently outperforms the much earlier \gru method.
Generally, our analyses point to certain underlying methodological issues, which were also observed in other application areas of applied machine learning. Similar observations regarding the competitiveness of established and often more simple approaches were made before, e.g., for the domains of information retrieval, time-series forecasting, and recommender systems, \cite{Yang:2019:CEH:3331184.3331340,Ferraridacremaetal2019,Makridakis2018,Armstrong:2009:IDA:1645953.1646031}, and it is important to note that these phenomena are not tied to deep learning approaches.

To help overcome some of these problems for the domain of session-based recommendation, we share our evaluation framework \textsc{\small{session-rec}} online\footnote{\url{https://github.com/rn5l/session-rec}}. The framework not only includes the algorithms that are compared in this paper, it also supports different evaluation procedures, implements a number of metrics, and provides pointers to the public datasets that were used in our experiments.

Since offline experiments cannot inform us about the quality of the recommendation as \emph{perceived} by users, we have furthermore conducted a user study. In this study, we compared heuristic methods with a neural approach and the recommendations produced by a commercial system (\textsc{\small{Spotify}}) in the context of an online radio station. The main outcomes of this study are that heuristic methods also lead to recommendations---playlists in this case---that are well accepted by users. The study furthermore sheds some light on the importance of other quality factors in the particular domain, i.e., the capability of an algorithm to help users discover new items.

The paper is organized as follows. Next, in Section \ref{sec:algorithms}, we provide an overview of the algorithms that were used in our experiments. Section \ref{sec:methodology} describes our offline evaluation methodology in more detail and Section \ref{sec:results} presents the outcomes of the experiments. In Section \ref{sec:user-study}, we report the results of our user study. Finally, we summarize our findings and their implications in Section \ref{sec:discussion}.

\section{Algorithms}
\label{sec:algorithms}
Algorithms of various types were proposed over the years for session-based recommendation problems. A detailed overview of the more general family of \emph{sequence-aware recommender systems}, where session-based ones are a part of, can be found in \cite{QuadranaetalCSUR2018}.
In the context of this work, we limit ourselves to a brief summary of parts of the historical development and how we selected algorithms for inclusion in our evaluations. 

\subsection{Historical Development and Algorithm Selection}
Nowadays, different forms of session-based recommendations can be found in practical applications. The recommendation of \emph{related items} for a given reference object can, for example, be seen as a basic and very typical form of session-based recommendations in practice. In such settings, the selection of the recommendations is usually based solely on the very last item viewed by the user. Common examples are the recommendation of additional articles on news web sites or recommendations of the form ``Customers who bought \dots also bought'' on e-commerce sites. Another common application scenario is the creation of automated playlists, e.g., on YouTube, Spotify, or Last.fm. Here, the system creates a virtually endless list of next-item recommendations based on some seed item and additional observations, e.g., skips or likes, while the media is played. These application domains---web page and news recommendation, e-commerce, music playlists---also represent the main driving scenarios in academic research.

For the recommendation of \emph{web pages} to visit, Mobasher et al. proposed one of the earliest session-based approaches based on frequent pattern mining in 2002 \cite{Mobasher2002}. In 2005, Shani et al. \citep{shani05mdp} investigated the use of an MDP-based (Markov Decision Process) approach for session-based recommendations in\emph{ e-commerce} and also demonstrated its value from a business perspective. Alternative technical approaches based on Markov processes were later on proposed in 2012 and 2013 for the \emph{news} domain in \cite{DBLP:conf/recsys/GarcinDF13} and \cite{DBLP:conf/webi/GarcinZFS12}.

An early approach to \emph{music playlist generation} was proposed in 2005 \cite{Ragno:2005:ISM:1101826.1101840}, where the selection of items was based on the similarity with a seed song. The music domain was however also very important for collaborative approaches. In 2012, the authors of \cite{hariri12context} used a session-based nearest-neighbors technique as part of their approach for playlist generation. This nearest-neighbors method and improved versions thereof later on turned out to be highly competitive with today's neural methods \cite{Ludewig2018}. More complex methods were also proposed for the music domain, e.g., an approach based on Latent Markov Embeddings \cite{Chen:2012:PPV:2339530.2339643} from 2012. 

Some novel technical proposals in the years 2014 and 2015 were based on a non-public \emph{e-commerce} dataset from a European fashion retailer and either used Markov processes and side information \cite{tavakol14fmdp} or a simple re-ranking scheme based on short-term intents \cite{Jannach2015}. More importantly, however, in the year 2015, the ACM RecSys conference hosted a challenge, where the problem was to predict if a consumer will make a purchase in a given session, and if so, to predict which item will be purchased. A corresponding dataset (YOOCHOOSE) was released by an industrial partner, which is very frequently used today for benchmarking session-based algorithms. Technically, the winning team used a two-stage classification approach and invested a lot of effort into feature engineering to make accurate predictions \cite{Romov:2015:RCE:2813448.2813510}.

In late 2015, Hidasi et al. \cite{Hidasi2016GRU} then published the probably first deep learning based method for session-based recommendation called \gru, a method which was continuously improved later on, e.g., in \cite{Hidasi:2018:RNN:3269206.3271761} or \cite{Tan2016GruPlus}. In their work, they also used the mentioned YOOCHOOSE dataset for evaluation, although with the slightly different optimization goal, i.e., to predict the immediate next item click event. As one of their baselines, they used an item-based nearest-neighbors technique. They found that their neural method is significantly better than this technique in terms of prediction accuracy. The proposal of their method and the booming interest in neural approaches subsequently led to a still ongoing wave of new proposals that apply deep learning approaches to session-based recommendation problems.

In this present work, we consider a selection of algorithms that reflects these historical developments. We consider basic algorithms based on item co-occurrences, sequential patterns and Markov processes as well as methods that implement session-based nearest-neighbors techniques. Looking at neural approaches, we benchmark the latest versions of \gru as well as five other methods that were published later and which state that they outperform at least the initial version of \gru to a significant extent.

Regarding the selected neural approaches, we limit ourselves to methods that do not use side information about the items in order to make our work easily reproducible and not dependent on such meta-data. Another constraint for the inclusion in our comparison is that the work was published in a 
major conferences, i.e., one that is rated A or A* according to the Australian CORE scheme. 
Finally, while in theory algorithms should be reproducible based on the technical descriptions in the paper, there are usually many small implementation details that can influence the outcome of the measurement. Therefore, like in \cite{Ferraridacremaetal2019}, we only considered approaches where the source code was available and could be integrated in our evaluation framework with reasonable effort.

\subsection{Considered Algorithms}

In total, we considered 12 algorithms in our comparison. Table \ref{tab:non-neural-baselines} provides an overview of the \emph{non-neural} methods. Table \ref{tab:neural-methods} correspondingly shows the neural methods considered in our analysis, ordered by their publication date.

\begin{table}[ht]
	\caption{Overview of the \emph{non-neural} methods compared in our analysis.}
	\label{tab:non-neural-baselines}
	\begin{tabularx}{1\textwidth}{@{}p{1.1cm}X@{}}
		\toprule
		\ar & This simple ``Association Rules'' method counts pairwise item co-occurrences in the training sessions. Recommendations for an ongoing session are generated by this method by returning those items that most frequently co-occurred with the last item of the current session in the past. For a formal definition, see \cite{Ludewig2018}.
		\\ \midrule
		\sr & This method called ``Sequential Rules'' was proposed in \cite{Ludewig2018}. It is similar to \ar in that it counts pairwise item co-occurrences in the training sessions. In addition to \ar, however, it considers the order of the items in a session and the distance between them using a decay function. The method often led to competitive results in particular in terms of the Mean Reciprocal Rank in the analysis in \cite{Ludewig2018}. \\ \midrule
		\sknn /\newline \vsknn & The analysis in \cite{JannachLudewig2017RecSys} showed that a simple session-based nearest-neighbors method similar to the one from \cite{Hariri2015} was competitive with the first version for \gru. Conceptually, the idea is to find past sessions that contain the same elements as the ongoing session. The recommendations are then based by selecting items that appeared in the most similar past session.
		\newline Since the sequence in which items are consumed in the ongoing user session might be of importance in the recommendation process, a number of ``sequential extensions'' to the \sknn method were proposed in \cite{Ludewig2018}. Here, the order of the items in a session proved to be helpful, both when calculating the similarities as well as in the item scoring process. Furthermore, according to \cite{Ludewig2018rsc} it can be beneficial to put more emphasis on less popular items by applying an Inverse-Document-Frequency(IDF) weighting scheme. In this paper, all those extensions are implemented in the \vsknn method.
		\\ \midrule
		\stan & This method called ``Sequence and
		Time Aware Neighborhood'' was presented at SIGIR '19 \cite{Garg:2019}. \stan is based on \sknn \cite{JannachLudewig2017RecSys}, but it additionally takes into account the following factors for making
		recommendations: i) the position of an item in the current session, ii) the recency of a past session w.r.t. to the current session, and iii) the position
		of a recommendable item in a neighboring session. Their results show that \stan significantly improves over \sknn, and is even
		comparable to recently proposed state-of-the-art deep learning approaches.
		\\ \midrule
		\vstan &  This method, which we propose in this present paper, combines the ideas from \stan and \vsknn in a single approach. It incorporates all three previously mentioned particularities of \stan, which already share some similarities with the \vsknn method. Furthermore, we add a sequence-aware item scoring procedure as well as the IDF weighting scheme from \vsknn.
		\\ \midrule
		\ct & This technique is based on Context Trees, which were originally proposed for lossless data compression.
		It is a non-parametric method and based on variable-order Markov models.
		The method was proposed in \cite{Mi2018ct}, where it showed promising results. \\ 
		\bottomrule
	\end{tabularx}
	\vspace{-10pt}
\end{table}

\begin{table}[ht]
	\caption{Overview of the \emph{neural} methods compared in our analysis.}
	\label{tab:neural-methods}
	\begin{tabularx}{1\textwidth}{@{}p{1.2cm}X@{}}
		\toprule
		\gru & \gru \cite{Hidasi2016GRU} was the first neural approach that employed RNNs  
		for session-based recommendation.
		This technique uses Gated Recurrent Units (GRU) \cite{DBLP:journals/corr/ChoMBB14} to deal with the vanishing gradient problem.
		The technique was later on improved using more effective loss functions \cite{Hidasi:2018:RNN:3269206.3271761}. \\ \midrule
		\narm & This model \cite{Li2017narm} extends \gru and improves its session modeling with the introduction of a hybrid encoder with an attention mechanism.
		The attention mechanism is in particular used to consider items that appeared earlier in the session and which are similar to the last clicked one.
		The recommendation scores for each candidate item are computed with a bilinear matching scheme based on the unified session representation.
		\\ \midrule
		\stamp & In contrast to \narm, this model \cite{Liu2018stamp} does not rely on an RNN.
		A short-term attention/memory priority model is proposed, which is
		(a) capable of capturing the users' general interests from the long-term memory of a session context, and which (b) also takes the users' most recent interests from the short-term memory into account.
		The users' general interests are captured by an external memory built from all the historical clicks in a session prefix (including the last click).
		The attention mechanism is built on top of the embedding of the last click that represents the user's current interests. \\ \midrule
		\nnet & This recent model \cite{Yuan2019nextitnet} also discards RNNs to model user sessions. In contrast to \stamp, convolutional neural networks are adopted with a few domain-specific enhancements.
		The generative model is designed to explicitly encode item inter-dependencies, which allows to directly estimate
		the distribution of the output sequence (rather than the desired
		item) over the raw item sequence.
		Moreover, to ease the optimization of the deep generative architecture, the authors propose to use residual networks to wrap convolutional layer(s) by residual block.  \\ \midrule
		\sgnn & This method \cite{DBLP:journals/corr/abs-1811-00855}
		models session sequences as graph structured data (i.e., directed graphs). Based on the session graph, \sgnn is capable of capturing transitions of items and generating item embedding vectors correspondingly, which are difficult to be revealed by
		conventional sequential methods like MC-based and RNN-based methods. With the help of item embedding vectors, \sgnn furthermore aims to construct reliable session representations from which the next-click item can be inferred.
		\\ \midrule
		\csrm & This method \cite{Wang:2019:CSR:3331184.3331210} is a hybrid framework that uses collaborative neighborhood information in session-based recommendations.
		\csrm consists of two parallel modules: an Inner Memory Encoder (IME) and an Outer Memory Encoder (OME). The IME models a user's own information in the current session with the help of Recurrent Neural Networks (RNNs) and an attention mechanism. The OME exploits collaborative information to better predict the intent of current sessions by investigating neighborhood sessions.
		Then, a fusion gating mechanism is used to selectively combine
		information from the IME and OME to obtain the final representation of the current session. Finally, \csrm obtains a recommendation score for each candidate item by computing a bi-linear match with the final representation of the current session. \\
		\bottomrule
	\end{tabularx}
	\vspace{-10pt}
\end{table}

Except for the \ct method, the non-neural methods from Table \ref{tab:non-neural-baselines} are conceptually very simple or almost trivial.
As mentioned above, this can lead to a number of potential practical advantages compared to more complex models, e.g., regarding online updates and explainability. 
From the perspective of the computational costs, the time needed to ``train'' the simple methods is often low, as this phase often reduces to counting item co-occurrences in the training data or to preparing some in-memory data structures. To make the nearest-neighbors technique scalable, we implemented the internal data structures and data sampling strategies proposed in \cite{JannachLudewig2017RecSys}. Specifically, we pre-process the training data to build fast in-memory look-up tables. These tables can then be used to almost immediately retrieve a set of potentially relevant neighbor sessions in the training data given an item in the test session. Furthermore, to speed up processing times, we sample only a fraction (e.g., 1,000)  of the most recent training sessions when we look for neighbors, as this proved effective in several application domains. 
In the end, the \ct method was the only one from the set of non-neural methods for which we encountered scalability issues in the form of memory consumption and prediction time when the set of recommendable items is huge.

Regarding alternative non-neural approaches, note that in our previous evaluation in \cite{Ludewig2018} only one neural method, but several other machine learning approaches were benchmarked. We do not include these alternative machine learning methods (\iknn, \fpmc, \mc, \smf, \bpr, \fism, \fossil)\footnote{\iknn: Item-based kNN \cite{Hidasi2016GRU}, \fpmc: Factorized Personalized Markov Chains \cite{Rendle2010}, \mc: Markov Chains \cite{markovchainnorris97}, \smf: Session-based Matrix Factorization \cite{Ludewig2018}, \bpr: Bayesian Personalized Ranking \cite{Rendle2009}, \fism: Factored Item Similarity Models \cite{kabbur13fism}, \fossil: FactOrized Sequential Prediction with Item SImilarity ModeLs \cite{he16fusing}.}. in our present analysis because the findings in \cite{Ludewig2018} showed that they either are generally not competitive or only lead to competitive results in few special cases. 

\begin{table}[ht]
	\caption{Overview of the baseline techniques that each neural session-based approach was originally compared to. The methods are ordered chronologically by the date of publication. The marks (\xmark) indicate which baselines were used in the comparison.}
	\label{tab:used-baselines-neural-approaches}
	\setlength{\tabcolsep}{4pt}
	\small
	\begin{tabularx}{1\textwidth}{@{}Xlcccccccc@{}}
		\toprule
		Method              & Publication    & \iknn & \sknn & \bpr & \fpmc & \gru & \narm & \stamp \\
		\midrule
		\gru   & ICLR (05/16)   & \xmark                    &                      & \xmark                   &                                           &                      &                      &                                             \\
		\igru  & RecSys (09/16) & \xmark                    &                      &                     &                      & \xmark                                         &                      &                       &                      \\
		\narm  & CIKM (11/17)   & \xmark                    &                      & \xmark                   & \xmark                    & \xmark                   &                      &                                             \\
		\stamp & KDD (08/18)    & \xmark                    &                      &                     & \xmark                    & \xmark                                       & \xmark                    &                                             \\
		\grutwo  & CIKM (10/18)   & \xmark                    &                      &                     &                      & \xmark                   &                      &                      \\
		\nnet  & WSDM (02/19)   &                      &                      &                     &                      &     \xmark                                   &                      &                                             \\
		\sgnn  & AAAI (02/19)   & \xmark                    &                      & \xmark                   & \xmark                    & \xmark                                         & \xmark                    & \xmark                                           \\
		\csrm  & SIGIR (07/19)  & \xmark                    & \xmark                    & \xmark                   & \xmark                    &       \xmark                     & \xmark                    &                                          \\
		\bottomrule
	\end{tabularx}
\end{table}

The development over time regarding the \emph{neural} approaches is summarized in Table \ref{tab:used-baselines-neural-approaches}. The table also indicates which baselines were used in the original papers. The analysis shows that \gru was considered as a baseline in all papers. Most papers refer to the original \gru publication from 2016 or an early improved version that was proposed shortly afterwards (which we term \igru here, see \cite{Tan2016GruPlus}). Most papers, however, do not refer to the improved version (\grutwo) discussed in \cite{Hidasi:2018:RNN:3269206.3271761}. Since the public code for \gru was constantly updated, we however assume that the authors ran benchmarks against the updated versions. \narm, as one of the earlier neural techniques, is the only neural method other than \gru that is considered quite frequently by more recent works.

The analysis of the used baselines furthermore showed that only one of the more recent papers proposing a neural method (\csrm) considers, i.e., \cite{Wang:2019:CSR:3331184.3331210}, session-based nearest-neighbors techniques as a baseline, even though their competitiveness was documented in a publication at the ACM Recommender Systems conference in 2017 \cite{JannachLudewig2017RecSys}. The authors of \cite{Wang:2019:CSR:3331184.3331210} (\csrm) however only consider the original proposal and not the improved versions from 2018 \cite{Ludewig2018}. The only other papers in our analysis, which consider session-based nearest-neighbors techniques as baselines, are about non-neural techniques (\ct and \stan). The paper proposing \stan furthermore is an exception in that since it considers quite a number of neural approaches (\grutwo, \stamp, \narm, \sgnn) in its comparison. 

\section{Evaluation Methodology}
\label{sec:methodology}
We benchmarked all methods under the same conditions, using the evaluation framework that we share online to ensure reproducibility of our results. 

\subsection{Datasets}
\label{subsec.datasets}
We considered eight datasets from two domains for our evaluation, e-commerce and music. Six of them are public and several of them were previously used to benchmark session-based recommendation algorithms. Table \ref{tab:datasets} briefly describes the datasets.

\begin{table} [h!t]
	\caption{Datasets used in the experiments}
	\label{tab:datasets}
	\begin{tabularx}{1\linewidth}{@{}lX@{}}
		\toprule
		\rsc & E-commerce dataset used in the 2015 ACM RecSys Challenge. \\
		\retailrocket & An e-commerce dataset from the company Retail Rocket. \\
		\diginetica & An e-commerce dataset shared by the company Diginetica. \\
		\zalando & A non-public dataset consisting of interaction logs from the European fashion retailer Zalando. \\
		\tmusic & Music listening logs obtained from Last.fm. \\
		\nowplaying & Music listening logs obtained from Twitter. \\
		\aotm & A public music dataset containing music playlists. \\
		\etracks & A private music dataset with hand-crafted playlists. \\
		\bottomrule
	\end{tabularx}
	\normalsize
\end{table}

We pre-processed the original datasets in a way that all sessions with only one interaction were removed. As done in previous works, we also removed 
items that appeared less than 5 times in the dataset. Multiple interactions with the same item in one session were kept in the data. While the repeated recommendation of an item does not lead to item discovery, such recommendations can still be helpful from a user's perspective, e.g., as reminders \cite{RepeatNet2019,Lerche2016,JannachLudewigLerche2017umuai}. Furthermore, we use an evaluation procedure where we run repeated measurements on several subsets (splits) of the original data, see Section \ref{subsec:evaluation-procedure}. The average characteristics of the subsets for each dataset are shown in Table \ref{tab:dataset-characteristics}. We share all datasets except \zalando and \etracks online.

\begin{table}[!ht]
	\setlength{\tabcolsep}{4pt}
	\centering
	\footnotesize
	\caption{Characteristics of the datasets. The values are averaged over all five splits.}
	\label{tab:dataset-characteristics}
	\begin{tabularx}{1\linewidth}{@{}Xrrrrrrrr@{}}
		\toprule
		Dataset            & \rsc &   \retailrocket & \diginetica & \zalando  & \tmusic  & \nowplaying & \aotm & \etracks \\ \midrule
		Actions            & 5.4M     & 210k      & 264k  & 4.5M    & 640k & 271k  & 307k & 1.5M \\
		Sessions           & 1.4M     & 60k       & 55k  & 365k     & 37k  & 27k & 22k & 132k \\
		Items              & 29k      & 32k       & 32k  & 189k     & 91k  & 75k  & 91k & 376k \\
		Days cov.       & 31       & 27        &  31   & 90      & 90 & 90 & 90 & 90 \\%
		\midrule
		Actions/Sess.   & 3.95  & 3.54  & 4.78  & 12.43  & 17.11 & 10.04 & 14.02 & 11.32 \\%
		Items/Sess.     & 3.17  & 2.56  & 4.01  & 8.39  & 14.47 & 9.38 & 14.01 & 11.31\\%
		Actions/Day       & 175k  & 8k    & 8.5k  & 50k  & 7k  & 3.0k  & 3.4k & 16.6k \\%
		Sessions/Day      & 44k   & 2.2k  & 1.7k  & 4k  & 300 & 243 & 243 & 1.4k \\%
		\bottomrule
	\end{tabularx}
\end{table}

\subsection{Evaluation Procedure and Metrics}
\label{subsec:evaluation-procedure}
\paragraph{Data and Splitting Approach.}
We apply the following procedure to create train-test splits. Since most datasets consist of time-ordered events, usual cross-validation procedures with the randomized allocation of events across data splits cannot be applied. Several authors only use one single time-ordered training-test split for their measurements. This, however, can lead to undesired random effects. We therefore rely on a protocol where we create five non-overlapping and contiguous subsets (splits) of the datasets. As done in previous works, we use the last \emph{n} days of each split for evaluation (testing) and the other days for training the models.\footnote{The number of days used for testing ($n$) was determined based on the characteristics of the dataset. We, for example, used the last day for the \rsc dataset, two for \retailrocket, five for the music datasets, and seven for \diginetica to ensure that train-test splits are comparable.} The reported measurements correspond to the averaged results obtained for each split.
The playlist datasets (\aotm and \etracks) are exceptions here as they do not have timestamps. For these datasets, we therefore randomly generated timestamps, which allows us to use the same procedure as for the other datasets.
Note that during the evaluation, we only considered items in the test split that appeared at least once in the training data.

\label{hyperparameters}
\paragraph{Hyperparameter Optimization.}
Proper hyperparameter tuning is essential when comparing machine learning approaches. We therefore tuned all hyperparameters for all methods and datasets in a systematic approach, using MRR@20 as an optimization target as done in previous works. Technically, we created subsets from the training data for validation. The size of the validation set was chosen in a way that it covered the same number of days that was used in the final test set. We applied a random hyperparameter optimization approach with 100 iterations as done in \cite{Hidasi:2018:RNN:3269206.3271761,Liu2018stamp,Li2017narm}. 
Since \narm and \csrm only have a smaller set of hyperparameters, we only had to do 50 iterations for these methods. 
Since the tuning process was particularly time-consuming for \sgnn and \nnet, we had to limit the number of iterations to 50 both for \sgnn on the \zalando dataset and for \nnet on the \rsc dataset.
The final hyperparameter values for each method and dataset can be found online, along with a description of the investigated ranges.

\paragraph{Accuracy Measures.}

For each session in the test set, we incrementally reveal one event of a session after the other, as was proposed in \cite{Hidasi2016GRU}\footnote{Note that the revealed items from a session can be used by an algorithm for the subsequent predictions, but the revealed interactions are not added to the training data.}. The task of the recommendation algorithm is to generate a prediction for the next event(s) in the session in the form of a ranked list of items. The resulting list can then be used to apply standard accuracy measures from information retrieval. The measurement can be done in two different ways.
\begin{itemize}
	\item As in \cite{Hidasi2016GRU} and other works, we can measure if the immediate next item is part of the resulting list and at which position it is ranked. The corresponding measures are the Hit Rate and the Mean Reciprocal Rank.
	\item In typical information retrieval scenarios, however, one is usually not interested in having one item right (e.g., the first search result), but in having as many predictions as possible right in a longer list that is displayed to the user. For session-based recommendation scenarios, this applies as well, as usually, e.g., on music and e-commerce sites, more than one recommendation is displayed. Therefore, we measure Precision and Recall in the usual way, by comparing the objects of the returned list with the entire remaining session, assuming that not only the immediate next item is relevant for the user. In addition to Precision and Recall, we also report the Mean Average Precision metric. 
\end{itemize}

The most common cut-off threshold in the literature is 20, probably because this was the chosen threshold by the authors of \gru \cite{Hidasi2016GRU}. We have made measurements for alternative list lengths as well, but will only report the results when using 20 as a list length in this paper. We report additional results for cut-off thresholds of 5 and 10 in an online appendix.\footnote{\url{https://rn5l.github.io/session-rec/umuai}}

\paragraph{Coverage and Popularity.}
Depending on the application domain, factors other than prediction accuracy might be relevant as well, including coverage, novelty, diversity, or serendipity \cite{Shani2011}. Since we do not have information about item characteristics, we focus on questions of coverage and novelty in this work.

With \emph{coverage}, we here refer to what is sometimes called ``aggregate diversity'' \cite{Adomavicius:2012:IAR:2197072.2197127}. Specifically, we measure the fraction of items of the catalog that ever appears in any top-n list presented to the users in the test set. This coverage measure in a way also evaluates the level of context adaptation, i.e., if an algorithm tends to recommend the same set of items to everyone or specifically varies the recommendations for a given session.

We approximate the \emph{novelty} level of an algorithm by measuring how popular the recommended items are on average. The underlying assumption is that recommending more unpopular items leads to higher novelty and discovery effects. Algorithms that mostly focus on the recommendation of popular items might be undesirable from a business perspective, e.g., when the goal is to leverage the potential of the long tail in e-commerce settings. Technically, we measure the \emph{popularity} level of an algorithm as follows. First, we compute min-max normalized popularity values of each item in the training set. Then, during evaluation, we compute the popularity level of an algorithm by determining the average popularity value of each item that appears in its top-n recommendation list.
Higher values correspondingly mean that an algorithm has a tendency to recommend rather popular items.

\paragraph{Running Times.}
Complex neural models can need substantial computational resources to be trained. Training a ``model'', i.e., calculating the statistics, for co-occurrence based approaches like \sr or \ar can, in contrast, be done very efficiently. For nearest-neighbors based approaches, actually no model is learned at all. Instead, some of our nearest-neighbors implementations need some time to create internal data structures that allow for efficient recommendation at prediction time. In the context of this paper, we will report running times for some selected datasets from both domains.

We executed all experiments on the same physical machine. The running times for the neural methods were determined using a GPU; the non-neural methods used a CPU. In theory, running times should be compared on the same hardware. Therefore, since the running times of the neural methods are much longer even when a GPU can be used, we can assume that the true difference in computational complexity is in fact even higher than we can see in our measurements.

\paragraph{Stability with Respect to New Data.}
In some application domains, e.g., news recommendation or e-commerce, new user-item interaction data can come in at a high rate. Since retraining the models to accommodate the new data can be costly, a desirable characteristic of an algorithm can be that the performance of the model does not degenerate too quickly before the retraining happens. To put it differently, it is desirable that the models do not overfit too much to the training data.

To investigate this particular form of model stability, we proceeded as follows. First, we trained a model on the training data $T_0$ of a given train-test split\footnote{We also optimized the hyperparameters on a subset of $T_0$ that was used as a validation set. The hyperparameters were kept constant for the remaining measurements.}. Then, we made measurements using two different protocols, which we term \emph{retraining} and \emph{no-retraining}, respectively.

\begin{itemize}
	\item In the \emph{retraining} configuration, we first evaluated the model that was trained on $T_0$ using the data of the first day of the test set. Then, we added this first day of the test set to $T_0$ and retrained the model on this extended dataset, which we name $T_1$. Then, we continued with the evaluation with the data from the second day of the test data, using the model trained on $T_1$. This process of adding more data to the training set, retraining the full model, and evaluating on the next day of the test set was done for all days of the test set except the last one.
	\item In the \emph{no-retraining} configuration, we also evaluated the performance day by day on the test data, but did not retrain the models, i.e., we used the model trained on $T_0$ for all days in the test data.
\end{itemize}
To enable a fair comparison in both configurations, we only considered items in the evaluation phase that appeared at least once in the original training data $T_0$.

Note that the absolute accuracy values for a given test day depends on the characteristics of the recorded data on that day. In some cases, the accuracy for the second test day can therefore even be higher than for the first test day, even if there was no retraining. An exact comparison of absolute values is therefore not too meaningful. However, we consider the \emph{relative} accuracy drop when using the initial model $T_0$ for a number of consecutive days as an indicator of the generalizability or stability of the learned models, provided that the investigated algorithms start from a comparable accuracy level.

\section{Results}
\label{sec:results}
In this section, we report the results of our offline evaluation. We will first focus on accuracy, then look at alternative quality measures, and finally discuss aspects of scalability and the stability of different models over time.

\subsection{Accuracy Results}

\begin{table}[htbp]
	\caption[Results for the e-commerce datasets]{Results for the e-commerce datasets, ordered by MAP@20. The best results for each metric are highlighted in bold font. The next best results for algorithms from the other category (either neural or non-neural) are underlined. Non-neural methods are marked with full circles, neural ones with empty ones.}
	\scriptsize
	\label{tab:results-ec}
	\begin{tabularx}{1\linewidth}{@{}Xrrr|rr|rr@{}}
		\toprule
		Metrics   & MAP@20     & P@20   & R@20   & HR@20   & MRR@20 & COV@20 & POP@20\\ \midrule
		\multicolumn{8}{c}{\retailrocket}                            \\ \midrule
		$\bullet$~\stan    &  \best{0.0285} & \best{0.0543} & \best{0.4748} &  \best{0.5938} &  \sign{0.3638} &  0.5929 &  0.0518 \\
		$\bullet$~\vstan  & 0.0284 & 0.0542 & 0.4741 & 0.5932	& 0.3636   & \scnd{0.5982}	& 0.0488     \\
		$\bullet$~\sknn    & 0.0283     & 0.0532 & 0.4707 & 0.5788  & 0.3370 & 0.5709	& 0.0540\\
		$\bullet$~\vsknn    & 0.0278     & 0.0531 & 0.4632 & 0.5745  & 0.3395 & 0.5562	& 0.0598\\
		$\circ$~\textit{\gru}
		& \scnd{0.0272} & \scnd{0.0502} & \scnd{0.4559} & \scnd{0.5669}  & 0.3237 & \sign{0.7973}	& \best{0.0347}\\
		$\circ$~\textit{\narm}      & 0.0270	& 0.0501 & 0.4526 & 0.5549 & 0.3196 & 0.6472	& 0.0569\\ 
		$\circ$~\textit{\csrm}   & 0.0252	& 0.0467	& 0.4246  & 0.5169  & 0.2955    & 0.6049	& 0.0496\\
		$\circ$~\textit{\sgnn}  & 0.0241	& 0.0441	& 0.4125  & 0.4998 &  \scnd{0.3252}  & 0.5521	& 0.0743\\
		$\circ$~\textit{\stamp}     & 0.0223	& 0.0420 & 0.3806  & 0.4620  & 0.2527 & 0.4865	& 0.0677\\ 
		$\bullet$~\ar        & 0.0205     & 0.0387 & 0.3533 & 0.4367  & 0.2407 & 0.5444	& 0.0527\\
		$\bullet$~\sr        & 0.0194     & 0.0362 & 0.3359 & 0.4174  & 0.2453 & 0.5185	& \scnd{0.0424}\\
		$\circ$~\textit{\mbox{\nnet}} & 0.0173     & 0.0320 & 0.3051 & 0.3779  & 0.2038 & 0.5737	& 0.0703\\
		$\bullet$~\ct        & 0.0162     & 0.0308 & 0.2902 & 0.3632  & 0.2305 & 0.4026	& 0.3740\\
		\midrule

		\multicolumn{8}{c}{\diginetica}                                 \\ \midrule
		$\bullet$~\sknn       & \best{0.0255}   & \best{0.0596} & 0.3715 & 0.4748 & 0.1714 & 0.8701	& 0.1026\\
		$\bullet$~\vstan  & 0.0252	& 0.0588	& \best{0.3723}	& \sign{0.4803}	& \sign{0.1837}    & 0.9384	& 0.0858    \\
		$\bullet$~\stan   &  0.0252 &  0.0589 &  0.3720 &  0.4800 &  0.1828 &  0.9161 &  0.0964 \\
		$\bullet$~\vsknn      & 0.0249   & 0.0584 & 0.3668 & 0.4729 & 0.1784 & \scnd{0.9419}	& 0.0840\\
		$\circ$~\textit{\gru}    & \scnd{0.0247} & \scnd{0.0577} & \scnd{0.3617} & \scnd{0.4639} & \scnd{0.1644} & \best{0.9498}	& \best{0.0567}\\
		$\circ$~\textit{\csrm}   & 0.0227	& 0.0544	& 0.3335	& 0.4258	& 0.1421   & 0.7337	& 0.0833 \\
		$\circ$~\textit{\narm}       & 0.0218 & 0.0528 & 0.3254 & 0.4188 & 0.1392 & 0.8696	& 0.0832\\
		$\circ$~\textit{\stamp}      & 0.0201 & 0.0489 & 0.3040 & 0.3917 & 0.1314 & 0.9188	& 0.0799\\
		$\bullet$~\ar         & 0.0189   & 0.0463 & 0.2872 & 0.3720  & 0.1280  & 0.8892	& 0.0863\\
		$\circ$~\textit{\sgnn}  & 0.0186	& 0.0451	& 0.2840	& 0.3638	& 0.1564  & 0.8593	& 0.1092 \\
		$\bullet$~\sr          & 0.0161	& 0.0401	& 0.2489	& 0.3277	& 0.1216 & 0.8736	& \scnd{0.0707}\\ 
		$\circ$~\textit{\mbox{\nnet}}  & 0.0149 & 0.0380 & 0.2416 & 0.2922 & 0.1424 & 0.7935	& 0.0947\\
		$\bullet$~\ct         & 0.0115 & 0.0294 & 0.1860 & 0.2494 & 0.1075 & 0.7554	& 0.4262\\
		\midrule

		\multicolumn{8}{c}{\zalando}                                 \\ \midrule
		$\bullet$~\vstan  & \best{0.0168}	& \sign{0.0777}	& \sign{0.2073}	& \sign{0.5362}	& 0.2488    & 0.5497	& \scnd{0.0664}    \\
		$\bullet$~\stan   &  0.0167 &  0.0774 &  0.2062 &  0.5328 &  0.2468 &  0.4918 &  0.0734 \\
		$\bullet$~\vsknn    & 0.0158   & 0.0740 & 0.1956 & 0.5162  & 0.2487 & \scnd{0.6246}	& 0.0680 \\
		$\bullet$~\sknn    & 0.0157     & 0.0738 & 0.1891 & 0.4352  & 0.1724 & 0.3316	& 0.0843 \\
		$\circ$~\textit{\sgnn}    &  \scnd{0.0146} &  \scnd{0.0700} &  \scnd{0.1823} &  0.4755 &  0.2804 &  0.3845 &  0.0865 \\
		$\circ$~\textit{\narm}      & 0.0144 & 0.0692 & 0.1795 & 0.4598 & 0.2248 & 0.3695	& 0.0837\\
		$\circ$~\textit{\csrm} &  0.0143 &  0.0695 &  0.1764 &  0.4500 &  0.2347 &  0.2767 &  0.0789     \\
		$\circ$~\textit{\gru}   & 0.0143     & 0.0666 & 0.1797 & \scnd{0.4925}  & \best{0.3069} & \best{0.6365}	& \sign{0.0403}\\
		$\bullet$~\sr        & 0.0136     & 0.0638 & 0.1739 & 0.4824  & \scnd{0.3043} & 0.5849	& 0.0696\\
		$\bullet$~\ar        & 0.0133     & 0.0631 & 0.1690 & 0.4665  & 0.2579 & 0.4672	& 0.0886\\
		$\bullet$~\ct     & 0.0118     & 0.0564 & 0.1573 & 0.4561  & 0.2993  & 0.4653	& 0.2564\\
		$\circ$~\textit{\stamp}     & 0.0104     & 0.0515 & 0.1359 & 0.3687  & 0.2065 & 0.2234	& 0.0868\\
		\midrule

		\multicolumn{8}{c}{\rsc}                                   \\ \midrule
		$\circ$~\textit{\narm}      & \best{0.0357}     & \best{0.0735} & \best{0.5109} & \scnd{0.6751}  & 0.3047 & 0.6399	& 0.0638\\
		$\circ$~\textit{\sgnn}  & 0.0351	& 0.0725	& 0.5060	& 0.6713	&  \best{0.3142}  & 0.5105	& 0.0720\\
		$\circ$~\textit{\mbox{\nnet}} & 0.0350 & 0.0722 & 0.5033 & 0.6691  & 0.3132 & 0.5295 & 0.0677 \\
		$\bullet$~\vstan  & \scnd{0.0350}	& \scnd{0.0718}	& \scnd{0.5080}	& \best{0.6761}	& 0.2943    & 0.6762	& \scnd{0.0634}\\
		$\circ$~\textit{\csrm}      &  0.0346 &  0.0714 &  0.4952 &  0.6566 &  0.2961 &  0.5929 &  0.0626 \\
		$\circ$~\textit{\stamp}     & 0.0344     & 0.0713 & 0.4979 & 0.6654  & 0.3033 & 0.5803	& 0.0655\\
		$\bullet$~\stan    &  0.0342 &  0.0701 &  0.4986 &  0.6656 &  0.2933 &  \scnd{0.6828} &  0.0773 \\
		$\bullet$~\vsknn    & 0.0341 & 0.0707 & 0.4937 & 0.6512  & 0.2872 & 0.6333	& 0.0777\\
		$\circ$~\textit{\gru}   & 0.0334     & 0.0682 & 0.4837 & 0.6480  & 0.2826 & \best{0.7482}	& \best{0.0294}\\
		$\bullet$~\sr        & 0.0332     & 0.0684 & 0.4853 & 0.6506  & 0.3010 & 0.6674	& 0.0716\\
		$\bullet$~\ar        & 0.0325     & 0.0673 & 0.4760 & 0.6361  & 0.2894 & 0.6297	& 0.0926\\
		$\bullet$~\sknn    & 0.0318     & 0.0657 & 0.4658 & 0.5996  & 0.2620 & 0.6099	& 0.0796\\
		$\bullet$~\ct        & 0.0316     & 0.0654 & 0.4710 & 0.6359  & \scnd{0.3072} & 0.6270	& 0.1446\\
		\bottomrule

	\end{tabularx}
	\vspace{-10pt}
\end{table}

\paragraph{E-Commerce Datasets.}

Table \ref{tab:results-ec} shows the results for the e-commerce datasets, ordered by the values obtained for the MAP@20 metric. The non-neural models are marked with full circles while the neural ones can be identified by empty ones. The highest value across all techniques is printed in bold; the highest value obtained by the other family of algorithms---neural or non-neural---is underlined. Stars indicate significant differences (p$<$0.05) according to a Kruskal–Wallis test between all the models and a Wilcoxon signed-rank test between the best-performing techniques from each category. The results for the individual datasets can be summarized as follows.

\begin{itemize}
	\item On the \retailrocket dataset, the nearest-neighbors methods consistently lead to the highest accuracy results on all the accuracy measures. Among the complex models, the best results were obtained by \gru on all the measures except for MRR, where \sgnn led to the best value. The results for \narm and \gru are almost identical on most measures.
	
	\item The results for the  \diginetica dataset are comparable, with the neighborhood methods leading to the best accuracy results. \gru is again the best method across the complex models on all the measures.
	
	\item For the \zalando dataset, the neighborhood methods dominate all accuracy measures, except for the MRR. Here, \gru is minimally better than the simple \sr method. Among the complex models, \gru achieves the best HR value, and the recent \sgnn method is the best one on the other accuracy measures.
	
	\item Only for the \rsc dataset, we can observe that a neural method (\narm) is able to slightly outperform our best simple baseline \vstan in terms of MAP, Precision and Recall. Interestingly, however, \narm is one of the earlier neural methods in this comparison. The best Hit Rate is achieved by \vstan; the best MRR by \sgnn. The differences between the best neural and non-neural methods are often tiny, in most cases around or less than 1\,\%.
\end{itemize}

Looking at the results across the different datasets, we can make the following additional observations.
\begin{itemize}
	\item Across all e-commerce datasets, the \vstan method proposed in this paper is, for most measures, the best neighborhood-based method. This suggests that it is reasonable to include it as a baseline in future performance comparisons.
	
	\item The ranking of the \emph{neural} methods varies largely across the datasets and does not follow the order in which the methods were proposed. Like for the non-neural methods, the specific ranking therefore seems to be strongly depending on the dataset characteristics. This makes it particularly difficult to judge the progress that is made when only one or two datasets are used for the evaluation.
	
	\item The results for the \rsc dataset are generally different from the other results. 
	Specifically, we found that some neural methods (\narm, \sgnn, \nnet) are competitive and sometimes even slightly outperform our baselines.
	Moreover, \stamp and \nnet are usually not among the top performers, but work well for this dataset.
	Unlike for other e-commerce datasets, \ct works particularly well for this dataset in terms of the MRR.
	Given these observations, it seems that the \rsc dataset has some unique characteristics that are different from the other e-commerce datasets. Therefore, it seems advisable to consider multiple datasets with different characteristics in future evaluations.
	
	\item We did not include measurements for \nnet, one of the most recent methods, for some of the datasets (\zalando, \tmusic, \etracks, \nowplaying), because our machines ran out of memory ($>$~32 GB). These datasets were either comparably large or had longer sessions on average.

\end{itemize}

\begin{table}[htbp]
	\caption[Results for the music domain datasets]{Results for the music domain datasets. The best results for each metric are highlighted in bold font. The next best results for algorithms from the other category (either neural or non-neural) are underlined. Again, non-neural methods are marked with a full circle, neural ones with an empty one.}
	\label{tab:results-music}
	\scriptsize
	\begin{tabularx}{1\linewidth}{@{}Xrrr|rr|rr@{}}
		\toprule
		Metrics   & MAP@20     & P@20   & R@20   & HR@20  & MRR@20 & COV@20 & POP@20 \\
		\midrule

		\multicolumn{8}{c}{\nowplaying}                             \\ \midrule
		$\bullet$~\mbox{\vsknn}    & \sign{0.0193}     & \best{0.0664} & \sign{0.1828} & 0.2534 & 0.0810 & 0.4661	& 0.0582\\
		$\bullet$~\sknn    & 0.0186     & 0.0655 & 0.1809 & 0.2450 & 0.0687 & 0.3150	& 0.0619\\
		$\bullet$~\stan   &  0.0175 &  0.0585 &  0.1696 &  0.2414 &  0.0871 &  \scnd{0.5128} &  0.0473 \\
		$\bullet$~\vstan  & 0.0174	& 0.0609	& 0.1795	& \sign{0.2597}	& 0.0853 & 0.4299	& 0.0505\\
		$\bullet$~\ar        & 0.0166     & 0.0564 & 0.1544 & 0.2076 & 0.0710 & 0.4531	& 0.0511\\
		$\bullet$~\sr        & 0.0133     & 0.0466 & 0.1366 & 0.2002 & 0.1052 & 0.4661	& \scnd{0.0383}\\
		$\circ$~\textit{\sgnn} & \scnd{0.0125}	& \scnd{0.0490}	& \scnd{0.1400}	& 0.2113	& 0.0935   & 0.3265	& 0.0576\\
		$\circ$~\textit{\narm}      & 0.0118  & 0.0463 & 0.1274 & 0.1849 & 0.0894 & 0.4715	& 0.0488\\
		$\circ$~\textit{\gru}   & 0.0116     & 0.0449 & 0.1361 & \scnd{0.2261} & \scnd{0.1076} & \sign{0.5795}	& \best{0.0286}\\
		$\circ$~\textit{\stamp}     & 0.0111
		& 0.0456	& 0.1244	& 0.1954	& 0.0921 & 0.2148	& 0.0714\\
		$\circ$~\textit{\csrm}    & 0.0095	& 0.0388	& 0.1065	& 0.1508	& 0.0594  & 0.2445	& 0.0494\\
		$\bullet$~\ct        & 0.0065     & 0.0287 & 0.0893 & 0.1679 & \best{0.1094} & 0.2714	& 0.2984\\
		\midrule
		
		\multicolumn{8}{c}{\tmusic}                                \\ \midrule
		$\bullet$~\mbox{\vsknn}    & \sign{0.0309}     & \sign{0.1090} & \sign{0.2347} & 0.3830 & 0.1162 & 0.3667	& 0.0485\\
		$\bullet$~\vstan  & 0.0296	& 0.1003	& 0.2306	& \sign{0.3904}	& 0.1564    & \scnd{0.4333}	& \scnd{0.0293}\\
		$\bullet$~\sknn    & 0.0290     & 0.1073 & 0.2217 & 0.3443 & 0.0898 & 0.1913	& 0.0574\\
		$\bullet$~\stan      &  0.0278 &  0.0949 &  0.2227 &  0.3830 &  0.1533 &  0.4315 &  0.0347 \\
		$\bullet$~\ar        & 0.0254     & 0.0886 & 0.1930 & 0.3088 & 0.0960 & 0.3524	& 0.0393\\
		$\bullet$~\sr        & 0.0240     & 0.0816 & 0.1937 & 0.3327 & 0.2410 & 0.4131	& 0.0317\\
		$\circ$~\textit{\narm}      & \scnd{0.0155} & \scnd{0.0675} & 0.1486 & 0.2956 & 0.1945 & 0.3858	& 0.0425\\
		$\circ$~\textit{\gru}   & 0.0150     & 0.0617 & \scnd{0.1529} & \scnd{0.3273} & \scnd{0.2369} & \best{0.4881}	& \best{0.0255}\\
		$\circ$~\textit{\csrm}    &  0.0118 &  0.0536 &  0.1236 &  0.2652 &  0.1503 &  0.2290 &  0.0390 \\
		$\circ$~\textit{\sgnn}  &  0.0108 &  0.0482 &  0.1151 &  0.2883 &  0.1894 &  0.3965 &  0.0412   \\
		$\circ$~\textit{\stamp}     & 0.0093     & 0.0411 & 0.0875 & 0.1539 & 0.0819 & 0.0852	& 0.0491\\
		$\bullet$~\ct        & 0.0058     & 0.0308 & 0.0885 & 0.2882 & \sign{0.2502} & 0.1932	& 0.4255\\
		\midrule

		\multicolumn{8}{c}{\aotm}                                   \\ \midrule
		$\bullet$~\sknn    & \sign{0.0037} & \sign{0.0139} & \sign{0.0390} & \sign{0.0417} & 0.0054 & 0.2937	& 0.1467\\
		$\bullet$~\mbox{\vsknn}    & 0.0032     & 0.0116 & 0.0312 & 0.0352 & 0.0057 & 0.5886	& 0.1199\\
		$\bullet$~\stan   &  0.0031 &  0.0126 &  0.0357 &  0.0402 &  0.0054 &  0.2979 &  0.1667 \\
		$\bullet$~\vstan  & 0.0024	& 0.0083	& 0.0231	& 0.0271	& 0.0060    & \sign{0.6907}	& \best{0.0566}\\
		$\bullet$~\ar        & 0.0018     & 0.0076 & 0.0200 & 0.0233 & 0.0059 & 0.5532	& 0.1049\\
		$\bullet$~\sr        & 0.0010     & 0.0047 & 0.0134 & 0.0186 & 0.0074 & 0.5669	& 0.0711\\
		$\circ$~\textit{\narm}      & \scnd{0.0009}     & \scnd{0.0050} & \scnd{0.0146} & \scnd{0.0202} & \scnd{0.0088} & 0.4816	& 0.1119\\
		$\bullet$~\ct        & 0.0006     & 0.0043 & 0.0126 & 0.0191 & \sign{0.0111} & 0.3357	& 0.4680\\
		$\circ$~\textit{\sgnn}  & 0.0006	& 0.0032	& 0.0096	& 0.0148	& 0.0082  & 0.4283	& 0.0812 \\
		$\circ$~\textit{\csrm}    & 0.0005	& 0.0040	& 0.0109	& 0.0100	& 0.0021    & 0.0056	& 0.6478\\
		$\circ$~\mbox{\textit{\nnet}} & 0.0004     & 0.0024 & 0.0071 & 0.0139 & 0.0065 & 0.4851	& 0.0960\\
		$\circ$~\textit{\stamp}     & 0.0003     & 0.0020 & 0.0063 & 0.0128 & \scnd{0.0088} & 0.5168	& 0.0872\\
		$\circ$~\textit{\gru}   & 0.0003     & 0.0020 & 0.0063 & 0.0130 & 0.0074 & \scnd{0.5898} & \scnd{0.0594}\\
		\midrule
		
		\multicolumn{8}{c}{\etracks}                                \\ \midrule
		$\bullet$~\sknn    & \sign{0.0024}	& \scnd{0.0129}	& \sign{0.0343}	& \sign{0.0377}	& 0.0054	& 0.2352	& 0.1622\\
		$\bullet$~\stan     &  0.0022 &  0.0119 &  0.0313 &  0.0357 &  0.0052 &  0.2971 &  0.1382 \\
		$\bullet$~\mbox{\vsknn}   & 0.0021	& 0.0110	& 0.0276	& 0.0312	& 0.0056	& 0.4572	& 0.1064\\
		$\bullet$~\vstan  & 0.0018	& 0.0086	& 0.0227	& 0.0265	& 0.0056	& \sign{0.5192}	& 0.0757\\
		$\circ$~\textit{\narm}    & \scnd{0.0018}	& \best{0.0131}	& \scnd{0.0311}	& \scnd{0.0345}	& \sign{0.0083}	& 0.0788	& 0.1589  \\
		$\circ$~\textit{\sgnn}  &  0.0017 &  0.0123 &  0.0301 &  0.0330 &  0.0077 &  0.0211 &  0.1833 \\
		$\bullet$~\ar        & 0.0016	& 0.0088	& 0.0219	& 0.0255	& \scnd{0.0071}	& 0.4529	& 0.0912\\
		$\circ$~\textit{\stamp}   & 0.0015	& 0.0114	& 0.0256	& 0.0272	& 0.0061	& 0.0405	& 0.1374  \\
		$\bullet$~\sr    &  0.0012 &  0.0067 &  0.0166 &  0.0201 &  \scnd{0.0071} &  0.4897 & \sign{0.0657} \\
		$\circ$~\textit{\csrm}   & 0.0011	& 0.0087	& 0.0189	& 0.0204	& 0.0048	& 0.0417	& 0.1587 \\
		$\circ$~\textit{\gru}  & 0.0007	& 0.0060	& 0.0132	& 0.0161	& 0.0051	& \scnd{0.2839}	& \scnd{0.0825} \\
		$\bullet$~\ct           &  0.0007 &  0.0054 &  0.0127 &  0.0170 &  \scnd{0.0071} &  0.2732 &  0.2685 \\
		\bottomrule
		
	\end{tabularx}
\end{table}

\paragraph{Music Domain.}
In Table \ref{tab:results-music} we present the results for the music datasets. In general, the observations are in line with what we observed for the e-commerce domain regarding the competitiveness of the simple methods.

\begin{itemize}
	
	\item Across all datasets excluding the \etracks dataset, the nearest-neighbors methods are consistently favorable in terms of Precision, Recall, MAP and the Hit Rate, and the \ct method leads to the best MRR. Moreover, the simple \sr technique often leads to very good MRR values.
	
	
	\item For \etracks dataset, the best Recall, MAP and the Hit Rate values are again achieved by neighborhood methods. The best Precision and the MRR values are, however, achieved by a neural method (\narm). 
	
	\item Again, no consistent ranking of the algorithms can be found across the datasets. In particular the neural approaches take largely varying positions in the rankings across the datasets. Generally, \narm seems to be a technique which performs consistently well on most datasets and measures. 
\end{itemize}

\subsection{Coverage and Popularity}
Table \ref{tab:results-ec} and Table \ref{tab:results-music} also contain information about the popularity bias of the individual algorithms and coverage information. Remember that we described in Section \ref{subsec:evaluation-procedure} how the numbers were calculated. From the results, we can identify the following trends regarding individual algorithms and the different algorithm families.

\paragraph{Popularity Bias.}
\begin{itemize}
	\item The \ct method is very different from all other methods in terms of its \emph{popularity bias}, which is much higher than for any other method.
	\item The \gru method, on the other hand, is the method that almost consistently recommends the most unpopular (or: novel) items to the users.
	\item The neighborhood-based methods are often in the middle. There are, however, also neural methods, in particular \sgnn, which seem to have a similar or sometimes even stronger popularity bias than the nearest-neighbors approaches. The assumption that nearest-neighbors methods are in general more focusing on popular items than neural methods can therefore not be confirmed through our experiments.
\end{itemize}

\paragraph{Coverage.}
\begin{itemize}
	\item In terms of \emph{coverage}, we found that \gru often leads to the highest values.
	\item The coverage of the neighborhood-based methods varies quite a lot, depending on the specific algorithm variant. In some configurations, their coverage is almost as high as for \gru, while in others the coverage can be low.
	\item The coverage values of the other neural methods also do not show a clear ranking, and they are often in the range of the neighborhood-based methods and sometimes even very low.
\end{itemize}

\subsection{Scalability}
We present selected results regarding the running times of the algorithms for two e-commerce datasets  and one music dataset in Table \ref{tab:running-times}. The reported times were measured for training and predicting for one data split. The numbers reported for predicting correspond to the average time needed to generate a recommendation for a session beginning in the test set. For this measurement, we used a workstation computer with an Intel Core i7-4790k processor and an Nvidia Geforce GTX 1080 Ti graphics card (Cuda 10.1/CuDNN 7.5).

\begin{table}[h!t]
	\caption{Running times for selected algorithms on two datasets.} 
	\label{tab:running-times}
	\begin{tabularx}{1\linewidth}{@{}Xrrrrrr@{}}
		\toprule
		& \multicolumn{3}{c}{Training (min)} & \multicolumn{3}{c}{Predicting (ms)} \\
		Algorithm & \rsc & \zalando & \etracks & \rsc & \zalando & \etracks \\
		\midrule

		$\circ$~\grutwo & 43.14               & 39.65                   & 12.54                   & 7.72                & 25.97                   & 278.23                  \\
		$\circ$~\stamp  & 32.51               & 133.17                  & 112.84                  & 14.94               & 55.45                   & 423.94                  \\
		$\circ$~\narm   & 225.82              & 797.72                  & 623.76                  & 7.83                & 25.00                   & 211.35                  \\
		$\circ$~\sgnn   & 827.37              & 1527.17                 & 482.46                  & 27.67               & 120.15                  & 797.97                  \\
		$\circ$~\csrm   & 156.89              & 203.15                  & 96.83                   & 24.98               & 66.93                   & 250.23                  \\
		$\circ$~\mbox{\nnet}   & 1577.40             & --                      & --                      & 8.98                & --                      & --                      \\ \hline
		$\bullet$~\ar     & 0.40                & 1.00                    & 0.34                    & 4.66                & 12.00                   & 105.43                  \\
		$\bullet$~\sr     & 0.41                & 0.53                    & 0.25                    & 4.66                & 11.77                   & 101.98                  \\
		$\bullet$~\sknn   & 0.18                & 0.13                    & 0.05                    & 37.82               & 27.77                   & 291.26                  \\
		$\bullet$~\vsknn  & 0.19                & 0.13                    & 0.05                    & 18.75               & 30.56                   & 278.51                  \\
		$\bullet$~\stan   & 0.18                & 0.20                    & 0.05                    & 36.78               & 33.26                   & 317.23                  \\
		$\bullet$~\vstan  & 0.18                & 0.13                    & 0.06                    & 21.33               & 55.58                   & 288.40                  \\
		$\bullet$~\ct     & 11.00               & 15.60                   & 4.35                    & 73.34               & 484.87                  & 1452.71                 \\
		
		\midrule
	\end{tabularx}
\end{table}

The results generally show that the computational complexity of neural methods is, as expected, much higher than for the non-neural approaches. In some cases, researchers therefore only use a smaller fraction of the original datasets, e.g., \nicefrac{1}{4} or \nicefrac{1}{64} of the \rsc dataset. Several algorithms---both neural ones and the \ct method---exhibit major scalability issues when the number of recommendable items increases. Furthermore, for the \nnet method, we found that it is consuming a lot of memory for some datasets, as mentioned above, leading to out-of-memory errors.

In some cases, like for \ct or \sgnn, not only the training time increases, but also the prediction times. In particular the prediction times can, however, be subject to strict time constraints in production settings. The prediction times for the nearest-neighbors methods are often slightly higher than those measured for methods like \gru, but usually lie within the time constraints of real-time recommendation (e.g., requiring about 30ms for one prediction for the \zalando dataset).

Since datasets in real-world environments can be even larger, this leaves us with questions regarding the practicability of some of the approaches. In general, even in case where a complex neural method would slightly outperform one of the more simple ones in an offline evaluation, it remains open if it is worth the effort to put such complex methods into production. For the \zalando dataset, for example, the best neural method (\sgnn) needs several orders of magnitude\footnote{The training time for \sgnn is 10.000 times higher than for \vstan.} more time to train than the best non-neural method \vstan, which also only needs half the time for recommending.

A final interesting observation is that there can be a large spread, i.e., in the range of an order of magnitude and more, between the running times of the neural methods. For example, the methods that use convolution (\nnet) or graph structures (\sgnn) often need much more time than other techniques like \gru or \narm. A detailed theoretical analysis of the computational complexity of the different algorithms and their underlying architectures is, however, beyond the scope of our present work, which compares the effectiveness and efficiency in an empirical way.

\subsection{Stability With Respect to New Data}
We report the stability results for the examined neural and non-neural algorithms 
in Table \ref{tab:stability_completed}. Given the computational requirements for this simulation-based analysis, which requires multiple full re-training phases, we selected one of the smaller datasets for each domain in this analysis, \diginetica and \nowplaying.%

We used two months of training data and 10 days of test data for both datasets, \diginetica and \nowplaying. The reported values show how much the accuracy results of each algorithm degrades (in percent), averaged across the test days when there is no daily retraining.

\begin{table}[h!t]
	\caption{Relative accuracy decrease (in percent) for the evaluated algorithms on two datasets, ordered by HR@20 for the \diginetica dataset. The best results for each metric are highlighted in bold font. The next best results from the other category (neural or non-neural) are underlined.}
	\label{tab:stability_completed}
	\begin{tabularx}{1\linewidth}{@{}Xrrrr@{}}
		\toprule
		& \multicolumn{2}{c}{\diginetica} & \multicolumn{2}{c}{\nowplaying} \\
		Metrics & HR@20 & MRR@20 & HR@20 & MRR@20  \\
		\midrule
		$\bullet$~\sknn              &  \underline{-1.90\,\%} &  \underline{-0.17\,\%} &  \underline{-23.42\,\%} &  \textbf{-14.29\,\%} \\
		$\bullet$~\vsknn             &  -2.28\,\% &  -0.64\,\% &   -27.20\,\% &  -14.36\,\% \\
		$\bullet$~\vstan             &  -2.53\,\% &  -0.64\,\% &  -28.53\,\% &  -28.22\,\% \\
		$\bullet$~\stan              &  -2.97\,\% &  -0.29\,\% &  -27.21\,\% &  -27.92\,\% \\
		$\bullet$~\ar                &  -4.83\,\% &  -5.33\,\% &  -29.76\,\% &  -33.94\,\% \\
		$\bullet$~\sr                &  -6.22\,\% &  -6.14\,\% &  -32.38\,\% &  -70.05\,\% \\
		$\bullet$~\ct                &  -7.98\,\% &  -6.94\,\% &  -50.49\,\% &  -85.97\,\% \\
		\midrule
		$\circ$~\narm             &  \textbf{-1.84\,\%} &    \textbf{0.30\,\%} &   -35.10\,\% &  -70.28\,\% \\
		$\circ$~\gru              &  -2.79\,\% &  -1.84\,\% &  -46.03\,\% &  -74.11\,\% \\
		$\circ$~\mbox{\nnet}             &  -3.75\,\% &  -4.69\,\% & - & - \\
		$\circ$~\sgnn             &  -3.76\,\% &  -2.14\,\% &  -46.05\,\% &  -75.74\,\% \\
		$\circ$~\csrm             &  -4.20\,\% &  -4.68\,\% &  \textbf{-17.84\,\%} &  \underline{-41.27\,\%} \\
		$\circ$~\stamp            &  -7.80\,\% &  -7.28\,\% &  -46.48\,\% &  -45.78\,\% \\
		\bottomrule
	\end{tabularx}
	
\end{table}

We can see from the results that the drop in accuracy without retraining can vary a lot across datasets (domains). For the \diginetica dataset, the decrease in performance ranges between 0 and 10 percent across the different algorithms and performance measures. The \nowplaying dataset from the music domain seems to be more short-lived, with more recent trends that have to be considered. Here, the decrease in performance ranges from about 15 to 50 percent in terms of HR and from about 15 to 85 percent in terms of MRR.\footnote{Generally, comparing the numbers across the datasets is not meaningful due to their different characteristics.}

Looking at the detailed results, we see that in both families of algorithms, i.e., neural and non-neural ones, some algorithms are much more stable than others when new data are added to a given dataset. For the 
non-neural approaches, we see that nearest-neighbor approaches are generally better than the other baselines techniques based on association rules or context trees.

Among the neural methods, \narm is the most stable one on the \diginetica dataset, but often falls behind the other deep learning methods on the \nowplaying dataset.\footnote{The experiments for \nnet could not be completed on this dataset because the method's resource requirements exceeded our computing capacities.} On this latter dataset, the \csrm method leads to the most stable results. In general, however, no clear pattern across the datasets can be found regarding the performance of the neural methods when new data comes in and no retraining is done.

Overall, given that the computational costs of training complex models can be high, it can be advisable to look at the stability of algorithms with respect to new data when choosing a method for production. According to our analysis, there can be strong differences across the algorithms. Furthermore, the nearest-neighbors methods appear to be quite stable in this comparison.

\section{Observations From a User Study}
\label{sec:user-study}

Offline evaluations, while predominant in the literature, can have certain limitations, in particular when it comes to the question of how the quality of the provided recommendations is \emph{perceived} by users. We therefore conducted a controlled experiment, in which we compared different algorithmic approaches for session-based recommendation in the context of an online radio station. In the following sections, we report the main insights of this experiment. While the study did not include all algorithms from our offline analysis, we consider it helpful to obtain a more comprehensive picture regarding performance of session-based recommenders. More details about the study can be found in \cite{ludewigjannach2019radio}.

\subsection{Research Questions and Study Setup}
\paragraph{Research Questions.}
Our offline analysis indicated that simple methods are often more competitive than the more complex ones. Our main research question therefore was how the recommendations generated by such simple methods are perceived by its users in different dimensions, in particular compared to recommendations by a complex method. Furthermore, we were interested how users perceive the recommendations of a commercial music streaming service, in our case \textsc{Spotify}, in the same situation.

\paragraph{Study Setup.}
An online music listening application in the form of an ``automated radio station'' was developed for the purpose of the study.
Similar to existing commercial services, users of the application could select a track they like (called a ``seed track''), based on which the application created a playlist of subsequent tracks. 

\begin{figure}[t!]
	\centering
	\includegraphics[width=0.88\textwidth]{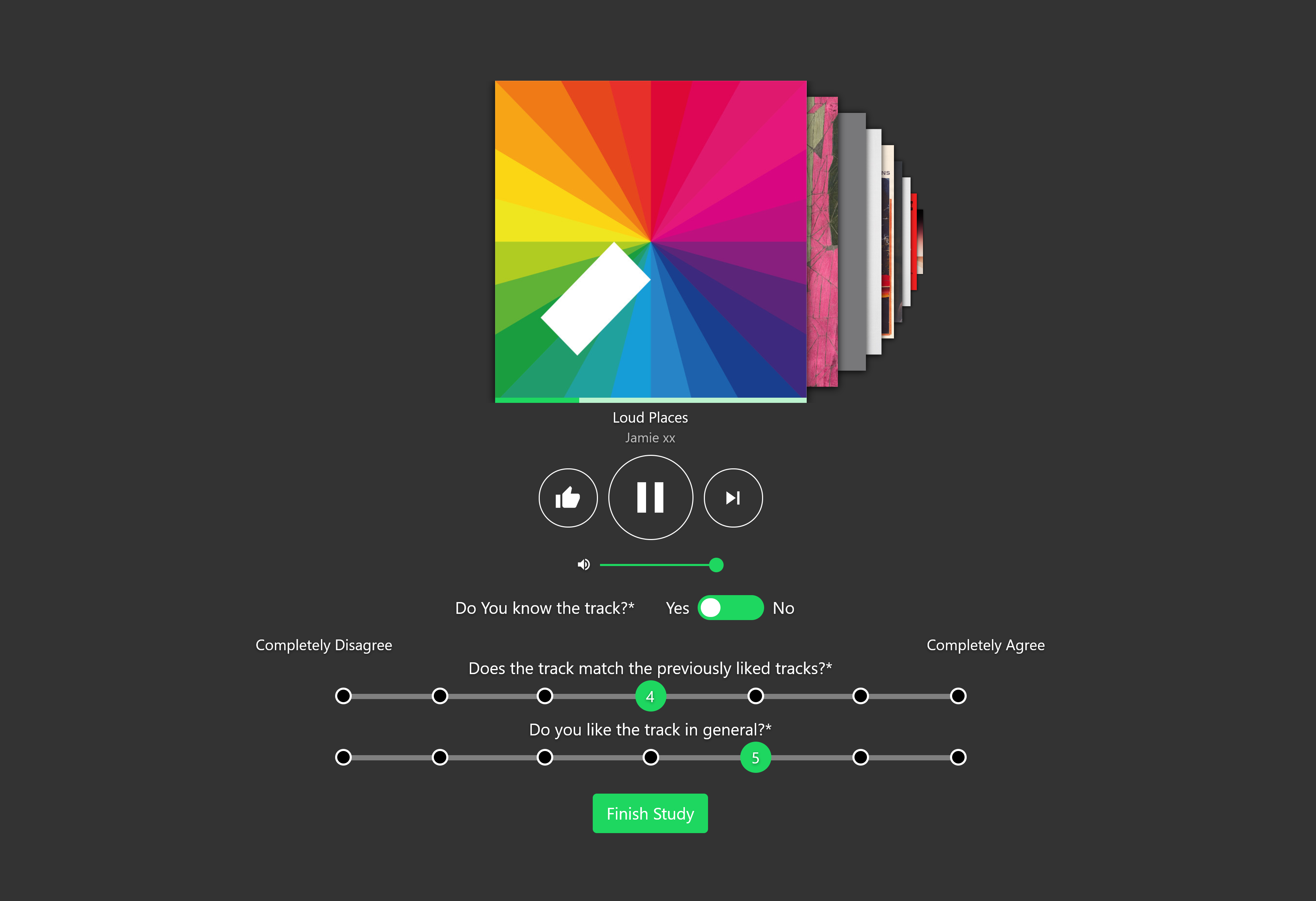}
	\caption{Track Rating Interface of the Application}
	\label{fig:radio2}
\end{figure}

The users could then listen to an excerpt of the next track and were asked to provide feedback about it as shown in in Figure \ref{fig:radio2}. Specifically, they were asked if \emph{(i)} if they already knew the track, \emph{(ii)} to what extent the track matched the previously played track, and \emph{(iii)} to what extent they liked the track (independent of the playlist). In addition, the participants could press a ``like'' button before skipping to the next track.  In case of such a \emph{like} action, the list of upcoming tracks was updated. Users were visually hinted that such an update takes place. This update of the playlist was performed for all methods including Spotify, i.e., in that case we re-fetched a new playlist through Spotify's API after each \emph{like} statement.



Once the participants had listened to and rated at least 15 tracks, they were forwarded to a post-task questionnaire. In this questionnaire, we asked the participants 11 questions about how they perceived the service, see also \cite{Pu:2011:UEF:2043932.2043962}. Specifically, the participants were asked to provide answers to the questions using seven-point Likert scale items, ranging from ``completely disagree'' to ``completely agree''. The questions, which include a twelfth question as an attention check, are listed in Table \ref{tab:quality-questions}. In the table, we group the questions according to the different quality dimensions they refer to, inspired by \cite{Pu:2011:UEF:2043932.2043962}. This grouping was not visible for participants in the online study.

\begin{table} [h!t]
	\caption{Questions about Users' Quality Perceptions.}
	\label{tab:quality-questions}
	\footnotesize
	\begin{tabularx}{1\linewidth}{@{}lX@{}}
		\toprule
		\multicolumn{2}{@{}l}{Question} \\
		\midrule
		\multicolumn{2}{@{}l}{\emph{Suitability of Tracks and Perceived Personalization}} \\
		\midrule
		Q1 & I liked the automatically generated radio station.  \\
		Q2 & The radio suited my general taste in music. \\
		Q3 & The tracks on the radio musically matched the track I selected in the beginning.  \\
		Q4 & The radio was tailored to my preferences the more positive feedback I gave. \\ \midrule
		\multicolumn{2}{@{}l}{\emph{Perceived  Diversity, Serendipity, and Familiarity}} \\
		\midrule
		Q5 & The radio was diversified in a good way. \\
		Q6 & The tracks on the radio surprised me.  \\
		Q7 & I discovered some unknown tracks that I liked in the process. \\ \midrule
		\multicolumn{2}{@{}l}{\emph{Attention Check}} \\ \midrule
		Q8 & I am participating in this study with care so I change this slider to two. \\ \midrule
		\multicolumn{2}{@{}l}{\emph{Intention to Reuse and to Recommend to Others}} \\   \midrule
		Q9 & I would listen to the same radio station based on that track again.  \\
		Q10 & I would use this system again, e.g., with a different first song. \\
		Q11 & I would recommend this radio station to a friend.  \\
		Q12 & I would recommend this system to a friend. \\ \bottomrule
	\end{tabularx}
	\normalsize
\end{table}

The study itself was based on a between-subjects design, where the treatments for each user group correspond to different algorithmic approaches to generate the recommendations. We included algorithms from different families in our study.

\begin{itemize}
	\item \ar: Association rules of length two, as described in Section \ref{sec:algorithms}. We included this method as a simple baseline.
	\item \cagh: Another relatively simple baseline, which recommends the greatest hits of artists similar to those liked in the current session. This music-specific method is often competitive in offline evaluations as well, see \cite{Bonnin:2014:AGM:2658850.2652481}.
	\item \sknn: The basic nearest-neighbors method described above. We took the simple variant as a representative for the family of such approaches, as it performed particularly well in the ACM RecSys 2018 challenge \cite{Ludewig2018rsc}.
	\item \gru: The RNN-based approach discussed above, used as a representative for neural methods. \narm would have been a stable alternative, but did not scale well for the used dataset.
	\item \spotify: Recommendations in this treatment group were retrieved in real time from Spotify's API.
\end{itemize}

We optimized and trained all models on the Million Playlist Dataset Million Playlist Dataset (MPD) \footnote{\url{https://recsys-challenge.spotify.com/}} provided by Spotify. We then recruited study participants using Amazon's Mechanical Turk crowdsourcing platform. After excluding participants who did not pass the attention checks, we ended up with \emph{N=250} participants, i.e., 50 for each treatment group, for which we were confident that they provided reliable feedback.

Most of the recruited participants (almost 80\%) were US-based. The most typical age range was between 25 and 34, with more than 50\% of the participants falling into this category. On average, the participants considered themselves to be music enthusiasts, with an average response of 5.75 (on the seven-point scale) to a corresponding survey question.
As usual, the participants received a compensation for their efforts through the crowdsourcing platform.

\subsection{User Study Outcomes}
The main observations can be summarized as follows.

\paragraph{Feedback the Listening Experience.}
Looking at the feedback that was observed during the listening session, we observed the following.

\begin{itemize}
	\item \emph{Number of Likes.} There were significant differences regarding the number of \emph{likes} we observed across the treatment groups. Recommendations by the simple \ar method received the highest number of likes (6.48), followed by \sknn (5.63), \cagh (5.38), \gru (5.36) and \spotify (4.48).
	
	\item \emph{Popularity of Tracks.} We found a clear correlation (\emph{r}=0.89) between the general popularity of a track in the MPD dataset and the number of likes in the study. The \ar and \cagh methods recommended, on average, the most popular tracks. The recommendations by \spotify and \gru were more oriented towards tracks with lower popularity.
	
	\item \emph{Track Familiarity.} There were also clear differences in terms of how many of the recommended tracks were already known by the users. The \cagh (10.83\,\%) and \sknn (10.13\,\%) methods recommended the largest number of known tracks. The \ar method, even though it recommended very popular tracks, led to much more unfamiliar recommendations (8.61\,\%). \gru was somewhere in the middle (9.30\,\%), and \spotify recommended the most novel tracks to users (7.00\,\%).
	
	\item \emph{Suitability of Track Continuations.} The continuations created by \sknn and \cagh were perceived to be the most suitable ones. The differences between \sknn and \ar, \gru, and \spotify were significant. The recommendations made by the \ar method were considered to match the playlist the least. This is not too surprising because the \ar method only considers the very last played track for the recommendation of subsequent tracks.
	
	\item \emph{Individual Track Ratings.} The differences regarding the individual ratings for each track ratings are generally small and not significant. Interestingly, the playlist-independent ratings for tracks recommended by the \ar method were the lowest ones, even though these recommendations received the highest number of likes. An analysis of the rating distribution shows that the \ar method often produces very bad recommendations, with a \emph{mode} value of 1 on the 1-7 rating scale.
\end{itemize}

\setlength{\tabcolsep}{3pt}
\begin{table} [h!t]
	\caption{Descriptive statistics and outcomes of the statistical significance tests for the post-task questionnaire. We report Mean, Standard Deviation (Mean $\pm$ Std), Median (Md), and Mode (Mo) for the Post-Task Questionnaire. We furthermore applied a Kruskal-Wallis test and subsequently a Mann-Whitney-U test when appropriate. Significant pairwise differences between the algorithms according to the Mann-Whitney-U test (p $<$ 0.05) are noted with \emph{k} for \sknn, \emph{c} for \cagh, \emph{g} for \gru, \emph{a} for \ar, and \emph{s} for Spotify.}
	\label{tab:user-study-statistics}
	\centering
	\footnotesize
	\begin{tabularx}{1\linewidth}{Xrrr@{\hspace{5pt}}rrr@{\hspace{5pt}}rrr}
		\toprule
		& \multicolumn{3}{c}{\knn}
		& \multicolumn{3}{c}{\cagh}                                                                                                          & \multicolumn{3}{c}{\gru}                                            \\
		& \multicolumn{1}{r}{Mean $\pm$ Std} & Md & Mo & \multicolumn{1}{r}{Mean $\pm$ Std} & Md & Mo & \multicolumn{1}{r}{Mean $\pm$ Std} & Md & Mo                   \\
		\midrule
		Q1 & $^{gas}$5.980 \tiny\textcolor{black!80}{$\pm$1.145} & 6& 7& $^{gas}$5.796 \tiny\textcolor{black!80}{$\pm$1.369} & 6& 6& 5.224 \tiny\textcolor{black!80}{$\pm$1.504} & 5& 5 \\
		Q2 & $^{a}$5.673 \tiny\textcolor{black!80}{$\pm$1.231} & 6& 6& $^{a}$5.735 \tiny\textcolor{black!80}{$\pm$1.483} & 6& 6& $^{a}$5.490 \tiny\textcolor{black!80}{$\pm$1.502} & 6& 7 \\
		Q3 & $^{gas}$5.673 \tiny\textcolor{black!80}{$\pm$1.281} & 6 & 7& $^{a}$5.286 \tiny\textcolor{black!80}{$\pm$1.646} & 6& 6& $^{a}$4.673 \tiny\textcolor{black!80}{$\pm$2.125} & 6& 6 \\
		Q4 & $^{a}$5.633 \tiny\textcolor{black!80}{$\pm$1.202} & 6& 6& $^{a}$5.510 \tiny\textcolor{black!80}{$\pm$1.697} & 6& 6& 5.224 \tiny\textcolor{black!80}{$\pm$1.531} & 5& 6 \\
		\midrule
		Q5 & 5.204 \tiny\textcolor{black!80}{$\pm$1.399} & 5& 5& 5.224 \tiny\textcolor{black!80}{$\pm$1.545} & 5& 5& 4.653 \tiny\textcolor{black!80}{$\pm$1.786} & 5& 4 \\
		Q6 & 3.878 \tiny\textcolor{black!80}{$\pm$1.589} & 4& 3& 3.755 \tiny\textcolor{black!80}{$\pm$1.774} & 4& 5& 4.000 \tiny\textcolor{black!80}{$\pm$1.720} & 4& 3 \\
		Q7 & 4.061 \tiny\textcolor{black!80}{$\pm$2.155} & 4& 7& 3.939 \tiny\textcolor{black!80}{$\pm$2.193} & 4& 1& 4.041 \tiny\textcolor{black!80}{$\pm$1.848} & 5& 5 \\
		\midrule
		Q9 & $^{as}$5.653 \tiny\textcolor{black!80}{$\pm$1.422} & 6& 6& $^{a}$5.347 \tiny\textcolor{black!80}{$\pm$1.809} & 6& 7& 5.082 \tiny\textcolor{black!80}{$\pm$1.730} & 5& 7 \\
		Q10 & $^{ga}$6.204 \tiny\textcolor{black!80}{$\pm$1.000} & 6& 7& $^{ga}$6.000 \tiny\textcolor{black!80}{$\pm$1.258} & 6& 7& 5.388 \tiny\textcolor{black!80}{$\pm$1.681} & 6& 7 \\
		Q11 & $^{a}$5.449 \tiny\textcolor{black!80}{$\pm$1.487} & 6& 7& $^{a}$5.408 \tiny\textcolor{black!80}{$\pm$1.790} & 6& 7& 4.959 \tiny\textcolor{black!80}{$\pm$1.744} & 5& 6 \\
		Q12 & $^{ga}$5.816 \tiny\textcolor{black!80}{$\pm$1.269} & 6& 6& $^{ga}$5.735 \tiny\textcolor{black!80}{$\pm$1.455} & 6& 7& 5.122 \tiny\textcolor{black!80}{$\pm$1.654} & 5& 5  \\
		\midrule
		&    \multicolumn{3}{c}{\ar}                                                                & \multicolumn{3}{c}{\spotify} & \multicolumn{3}{c}{}                                                            \\
		\midrule
		Q1 & 4.776 \tiny\textcolor{black!80}{$\pm$1.598} & 5& 3 & $^{a}$5.367 \tiny\textcolor{black!80}{$\pm$1.453} & 6& 6 & & &\\
		Q2 & 4.735 \tiny\textcolor{black!80}{$\pm$1.765} & 5& 3 & 5.306 \tiny\textcolor{black!80}{$\pm$1.475} & 5& 5 & & & \\
		Q3 & 4.245 \tiny\textcolor{black!80}{$\pm$1.843} & 4& 2 & 4.980 \tiny\textcolor{black!80}{$\pm$1.548} & 5& 5 & & & \\
		Q4 & 5.082 \tiny\textcolor{black!80}{$\pm$1.205} & 5& 4 & $^{a}$5.592 \tiny\textcolor{black!80}{$\pm$1.273} & 6& 7 & & & \\
		\midrule
		Q5 & 4.633 \tiny\textcolor{black!80}{$\pm$1.603} & 5& 3 & 4.959 \tiny\textcolor{black!80}{$\pm$1.707} & 5& 5 & & & \\
		Q6 & 4.204 \tiny\textcolor{black!80}{$\pm$1.384} & 5& 5 & 4.286 \tiny\textcolor{black!80}{$\pm$1.620} & 4& 3 & & & \\
		Q7 & 4.286 \tiny\textcolor{black!80}{$\pm$2.189} & 6& 6 & $^{kcga}$5.224 \tiny\textcolor{black!80}{$\pm$1.476} & 5& 5 & & & \\
		\midrule
		Q9 & 4.755 \tiny\textcolor{black!80}{$\pm$1.362} & 4& 4 & $^{a}$5.224 \tiny\textcolor{black!80}{$\pm$1.476} & 5& 6 & & & \\
		Q10 & 5.245 \tiny\textcolor{black!80}{$\pm$1.465} & 5& 4 & $^{ga}$6.041 \tiny\textcolor{black!80}{$\pm$1.274} & 6& 7 & & & \\
		Q11 & 4.490 \tiny\textcolor{black!80}{$\pm$1.647} & 4& 3 & $^{a}$5.265 \tiny\textcolor{black!80}{$\pm$1.524} & 5& 7 & & & \\
		Q12 & 4.796 \tiny\textcolor{black!80}{$\pm$1.720} & 5& 3 & $^{a}$5.551 \tiny\textcolor{black!80}{$\pm$1.473} & 6& 7  & & & \\
		\bottomrule
	\end{tabularx}
\end{table}

\paragraph{Post-Task Questionnaire.} The detailed statistics of the answers to the post-task questionnaire are shown in Table \ref{tab:user-study-statistics}. The analysis of the data revealed the following aspects:
\begin{itemize}
	\item Q1: The radio station based on \sknn was significantly more liked than the stations that used \gru, \ar, and \spotify.
	\item Q2: All radio stations matched the users general taste quite well, with median values between 5 and 6 on a seven-point scale. Only the station based on the \ar method received a significantly lower rating than the others.
	\item Q3: The \sknn method was found to perform significantly better than \ar and \gru with respect to identifying tracks that musically match the seed track.
	\item Q4: The adaptation of the playlist based on the like statements was considered good for all radio stations. Again, the feedback for the \ar method was significantly lower than for the other methods.
	\item Q5 and Q6: No significant differences were found regarding the surprise level of the different recommendation strategies.
	\item Q7: Regarding the capability of recommending unknown tracks that the users liked, the recommendations by \spotify were perceived to be much better than for the other methods, with significant differences compared to all other methods.
	\item Q9 to Q12: The best performing methods in terms of the intention to reuse and the intention to recommend the radio station to others were \sknn, \cagh, and \spotify. \gru and \ar were slightly worse, sometimes with differences that were statistically significant.
\end{itemize}

Overall, the study confirmed that methods like \sknn do not only perform well in an offline evaluation, but are also able, according to our study, to generate recommendations that are well perceived in different dimensions by the users. The study also revealed a number of additional insights.

First, we found that optimizing for \emph{like} statements can be misleading. The \ar method received the highest number of likes, but was consistently worse than other techniques in almost all other dimensions. Apparently, this was caused by the fact that the \ar method made a number of bad recommendations; see also \cite{CHAU2013180} for an analysis of the effects on bad recommendations in the music domain.

Second, it turned out that \emph{discovery support} seems to be an important factor in this particular application domain. While the recommendations of \spotify were slightly less appreciated than those by \sknn, we found no difference in terms of the user's intention to reuse the system or to recommend it to friends. We hypothesize that the better discovery support of \spotify's recommendations was an important factor for this phenomenon. This observation points to the importance of considering multiple potential quality factors when comparing systems.

\section{Research Limitations}
Our work does not come without limitations, both regarding the offline evaluations and the user study.

\paragraph{Potential Data Biases.}
One general problem of offline evaluations based on historical data is that we often know very little about the circumstances and environment in which the data was collected. For the e-commerce datasets, for example, what we see as interactions in the log can be at least partially the result of the recommender system that was in place during the time of data collection, or it can simply be the result of how certain items or categories were promoted in the online shop. 
For the music datasets, and in particular for data obtained from Last.fm (\tmusic), it might be that the logs to some extent reflect what the Last.fm radio station functionality was playing automatically given a seed track. Well-performing algorithms, i.e., those that predict the next items in the log accurately, might therefore be the ones that are able to ``reconstruct'' the logic of an existing recommender in some ways. 
The results of such a biased offline evaluation might therefore not fully reflect the effectiveness of a system.

Over the years, a number of approaches were proposed to deal with such problems, e.g., by using evaluation measures that take biases into account or by trying to ``de-bias'' the datasets \cite{Steck:2010:TTR:1835804.1835895,Debiased2019}. In particular in the context of reinforcement learning and bandit-based approaches, a number of research proposals were made for unbiased offline evaluation protocols to obtain more realistic performance estimates from log data, see \cite{Li:2011:UOE:1935826.1935878} for an early work. The analysis or consideration of such biases was, however, not in the scope of the work, which aimed at the comparison of different existing algorithms using standard evaluation protocols. While the outcomes of these analyses (and of the original works) maybe therefore suffer from potential biases, the conducted user study provided us with strong indications that the generated recommendations were also liked by users.

\paragraph{Empirical Nature of the Work.}
Generally, our work---like the papers that proposed the analyzed neural models---is mainly an empirical one in terms of the research approach. Algorithmic papers that propose new models in many cases do not start with a theoretical model, but probably more often with an intuition of what kind of signals there could be in the data. In case performance increases are found when using model that is designed to capture these signals, a common approach in that context is to use ablation studies to determine, again empirically, to what extent certain parts of the network architecture contribute to the overall performance. In the context of our comparative work, in contrast, it would be interesting to understand why even computationally very complex models are \emph{not} consistently performing better than the more simple models. Possible explanations could be that some underlying assumptions do not hold for the majority of the datasets. In some domains, the sequential ordering of the events, as captured by RNNs, might for example not be very important. Another problem could lie in a certain tendency of overfitting of the complex models, even when the hyperparameters are optimized on a held-out validation set. A detailed analytical investigation of the potential reasons why each of the six complex models in our comparison does \emph{not} perform consistently better than the more simple ones, however, lies beyond the scope of this present work and is left for future work.

\paragraph{User Study Limitations.}
Finally, the user study discussed in Section \ref{sec:user-study}---like most studies of that type---has certain limitations as well. Typical issues that apply also to our study are questions related the representativeness of the user population. Furthermore, while we developed a realistic and fully functional online radio station, the setting remains artificial and users were paid for their participation. The attention checks and the statistics of how users interacted with the system however make us confident that the majority of the participants completed the task with care and that the results are reliable. Another potential limitation of our study design is that we used one single item for each of the investigated quality dimensions in the post-task questionnaire. Since we mainly used established questions from the literature, e.g., from \cite{Pu:2011:UEF:2043932.2043962}, the associated risks are low.

\section{Conclusions and Ways Forward}
\label{sec:discussion}
Our work reveals that despite a continuous stream of papers that propose new neural approaches for session-based recommendation, the progress in the field seems still limited. According to our evaluations, today's deep learning techniques are in many cases not outperforming much simpler heuristic methods. Overall, this indicates that there still is a huge potential for more effective neural recommendation methods in the future in this area.

In a related analysis of deep learning techniques for recommender systems \cite{Ferraridacremaetal2019,Ferraridacrema2019troubling}, the authors found that different factors contribute to what they call \emph{phantom progress}. One first problem is related to the reproducibility of the reported results. They found that in less than a third of the investigated papers, the code was made available to other researchers. The problem also exists to some extent for session-based recommendation approaches. To further increase the level of reproducibility, we share our evaluation framework publicly, so that other researchers can easily benchmark their own methods with a comprehensive set of neural and non-neural approaches on different datasets.

Through sharing our evaluation framework, we hope to also address other methodological and procedural issues mentioned in \cite{Ferraridacremaetal2019} that can make the comparison of algorithms unreliable or inconclusive. Regarding methodological issues, we for example found works that determined the optimal number of training epochs on the test set and furthermore determined the best Hit Rate and MRR values across different optimization epochs. Regarding procedural issues, we found that while researchers seemingly rely on the same datasets as previous works, they sometimes apply different data pre-processing strategies. Furthermore, the choice of the baselines can make the results inconclusive. Most investigated works do not consider the \sknn method and its variants as a baseline. Some works only compare variants of one method and include a non-neural, but not necessarily strong other baseline. In many cases, little is also said about the optimization of the hyperparameters of the baselines. The \textsc{\small{session-rec}} framework used in our evaluation should help to avoid these problems, as it contains all the code for data pre-processing, evaluation, and hyperparameter optimization. Such frameworks are generally important to ensure replicability and reproducibility of research results \cite{CobaZanker2017}. Furthermore, sharing the framework allows other researchers to inspect the exact details of how the algorithms are implemented and evaluated, which is important as no \emph{de-facto} standards exists in the literature, which can sometimes lead to inconclusive and inconsistent results \cite{SaidBellogin2014}.

Moreover, also from a methodological perspective, our analyses indicated that optimizing solely for accuracy can be insufficient also for session-based recommendation scenarios. Depending on the application domain, other quality factors such as coverage, diversity, or novelty should be considered besides efficiency, because they can be crucial for the adoption and success of the recommendation service. Given the insights from our controlled experiment, we furthermore argue that more user studies and field tests are necessary to understand the characteristics of successful recommendations in a given application domain.

Looking at future directions, in particular methods that leverage side information about users and items seem to represent a promising way forward, see \cite{gabrieljannach2019,deSouzaPereiraMoreira2018,Huang2018,Hidasi:2016:PRN:2959100.2959167}. In \cite{Hidasi:2016:PRN:2959100.2959167}, the authors for example use a parallel RNN architecture to incorporate image and text information in the session modeling process. In \cite{deSouzaPereiraMoreira2018} and \cite{gabrieljannach2019}, both item information and user context information are combined in a neural architecture for news recommendation. The authors of \cite{Huang2018}, finally, combine RNNs with Key-Value memory networks to build a hybrid system that integrates information about item attributes in the sequential recommendation process.

A main challenge when trying to analyze and compare such methods under identical conditions, as was the goal of our present work, is that these works rely on largely different and often specific datasets, e.g., containing image information, or are optimized for a specific problem setting, e.g., cold-start situations in the news domain. An important direction for future work therefore lies in analyzing to what extent the benefits of such hybrid architectures generalize beyond individual application domains.
\color{black}

\section*{Acknowledgement}
We thank Liliana Ardissono for her valuable feedback on the paper.

\bibliographystyle{plain}
\bibliography{lit}

\end{document}